# A current-voltage model for double Schottky barrier devices

*Alessandro Grillo and Antonio Di Bartolomeo*[*]

Alessandro Grillo, Author 1,
Physics Department "E. R. Caianiello" and Interdepartmental Center "Nanomates",
University of Salerno, via Giovanni Paolo II n. 132, Fisciano 84084, Italy
E-mail: agrillo@unisa.it

Prof. Antonio Di Bartolomeo, Author 2
Physics Department "E. R. Caianiello" and Interdepartmental Center "Nanomates",
University of Salerno, via Giovanni Paolo II n. 132, Fisciano 84084, Italy
E-mail: adibartolomeo@unisa.it

Keywords: Schottky Barrier, Current-Voltage characteristic, Diode, Transition Metal Dichalcogenide.

Schottky barriers are often formed at the semiconductor/metal contacts and affect the electrical behaviour of semiconductor devices. In particular, Schottky barriers have been playing a major role in the investigation of the electrical properties of mono and two-dimensional nanostructured materials, although their impact on the current-voltage characteristics has been frequently neglected or misunderstood. In this work, we propose a single equation to describe the current-voltage characteristics of two-terminal semiconductor devices with Schottky contacts. We apply the equation to numerically simulate the electrical behaviour for both ideal and non-ideal Schottky barriers. The proposed model can be used to directly estimate the Schottky barrier height and the ideality factor. We apply it to perfectly reproduce the experimental current-voltage characteristics of ultrathin molybdenum disulphide or tungsten diselenide nanosheets and tungsten disulphide nanotubes. The model constitutes a useful tool for the analysis and the extraction of relevant transport parameters in any two-terminal device with Schottky contacts.

## 1. Introduction

Current-voltage (I-V) measurements are an efficient way for evaluating the quality of metal contacts and for the extraction of fundamental parameters such as the Schottky barrier (SB) height and the contact resistance. To study the I-V characteristic of two terminal devices with SBs, the classic equation of the Schottky diode is often used. This equation considers that charge carriers cross an energy barrier by thermionic emission. In this approach, it is implicitly assumed that there is a Schottky junction at the forced contact while the other contact is ohmic. Moreover, in such an approach, only the dominant current, occurring either at positive or negative voltages, is commonly used to evaluate the barrier parameters. Nevertheless, in most devices both contacts have a rectifying nature and the simplistic assumption of a single SB leads to the wrong results, such as very large ideality factors ($n \gg 1$) or unrealistic SB height and Richardson constant.

Oldham and Milnes highlighted the need to consider two barriers to describe the electric conduction based on thermionic emission in n-n heterojunctions.[1] Then, Ramirez et al.[2] and Nagano et al.[3] applied the model to study the electric transport in $SnO_2$ nanowires and in $C_{60}$ films with Au contacts. In both works, the authors did not consider that barriers with different heights could form at the contacts. Such a possibility might easily occur as the SB height is related to interfacial chemistry and local defects. Nouchi proposed a single equation[4] to study the Schottky barrier of a device formed by a metal-semiconductor junction, in which a second electrode necessary to perform the electrical measurement forms a metal-semiconductor-metal structure. This model does not consider the possibility that the two Schottky barriers formed with the contacts may have different ideality factors and requires the use of the first and second derivative of the I-V characteristic of the device that can limit its applicability. A model with two different barriers at the contacts was proposed by Chiquito et al. who limited their study to the analysis of metal/$SnO_2$/metal devices.[5] A model based on SBs with different heights was used by Di Bartolomeo et al. to account for the rectifying electrical characteristics observed in field effect transistors with ultrathin transition-metal dichalcogenide channel[6,7] or in metal/insulator/semiconductor structures.[8]

In this work, we apply a simple model to describe the conduction in a two-terminal device in which two rectifying contacts are present. The model allows extracting both SB heights simultaneously. First, we use the model to make numerical simulations of the I-V curves for different combination of junction parameters such as barrier height, area and ideality factor.[9] Then, we test the model fitting the symmetric or asymmetric I-V curves of two-terminal nanodevices made of molybdenum disulphide (MoS$_2$) or tungsten diselenide (WSe$_2$) nanosheets and tungsten disulphide (WS$_2$) nanotubes. MoS$_2$, WSe$_2$ and WS$_2$ are major exponents of the transition metal dichalcogenides, a family of layered materials with promising technological applications.[6,7,10–12] We highlight that, although demonstrated for nanostructured semiconductors, the model can be applied to any devices with two SBs.

## 2. Method

The SB height is defined as:

$$\phi_{SB-n} = \phi_m - \chi_s \qquad \phi_{SB-p} = E_g + \chi_s - \phi_m \qquad (1)$$

where $\phi_{SB-n}$ and $\phi_{SB-p}$ are the SB heights for electron and hole injection, respectively, $\phi_m$ is the metal work function, $\chi_s$ is the semiconductor electron affinity and $E_g$ is the bandgap of the semiconductor, as shown in **Figure 1a**. According to **Equations 1**, a low work function metal with the Fermi level close or above the bottom of the conduction band of the semiconducting material facilitates electron injection whereas a high work function metal with the Fermi level close or below the valence band of the material favours hole injection. However, equations 1 give only qualitative indications of the SB heights and are rarely in quantitative agreement with the experimental data. This is mainly because the metal Fermi level can be pinned at a semiconductor interface state, $\phi_{IS}$, if a high-density of interface states is formed within the semiconductor bandgap during the fabrication process, as shown in **Figure 1b**. In this case, the equation for $\phi_{SB-n}$ is replaced by:[13]

$$\phi_{SB-n} = (S \cdot \phi_m - \chi_s) + (1-S)\phi_{IS} \qquad (2)$$

where $S = \frac{\partial \phi_{SB-n}}{\partial \phi_M}$ is the Schottky pinning factor (a similar expression is used for $\phi_{SB-p}$). If $S = 0$, the pinning is maximum, the SB turns out to be independent of the metal work function and the so-called Bardeen limit is reached. Conversely, when $S = 1$, the Bardeen limit converges to the Schottky limit, represented by eq. (1).

Since the evaluation of the Schottky pinning factor and the semiconductor interface state energy, $\phi_{IS}$, is complicated by variability of the fabrication process, the extraction of the SB height by means of electrical measurements is one of the most efficient techniques of investigation.

Usually, when a semiconductor material is measured in standard two-probe configuration, the grounded contact is assumed ohmic while the other contact can form a Schottky junction regulated by a potential barrier. This assumption is unrealistic because Schottky junctions can be formed at both contacts. Hence, to approach the problem in a more general way, we assume that there are SBs at both metal contacts with the semiconductor.

If no external potential is applied, the band diagram of the device with two-Schottky contacts is shown in **Figure 1c** in which the two barriers are designated as $\phi_{B01}$ and $\phi_{B02}$, respectively. When a positive potential $V$ (bias) is applied to the left contact, the bands are bent as shown in **Figure 1d**. We indicate with $V_{1,2}$ and $I_{1,2}$ the potential drop and the currents at the contacts 1 and 2, respectively.

We note that the semiconductor with SBs at both contacts can be modelled with two back-to-back Schottky diodes separated by a series resistance. When a sufficiently high external voltage is applied, whether positive or negative, one Schottky junction is forward-biased while the other one is reverse-biased. The reverse saturation current of the reverse-biased diode always limits the current.

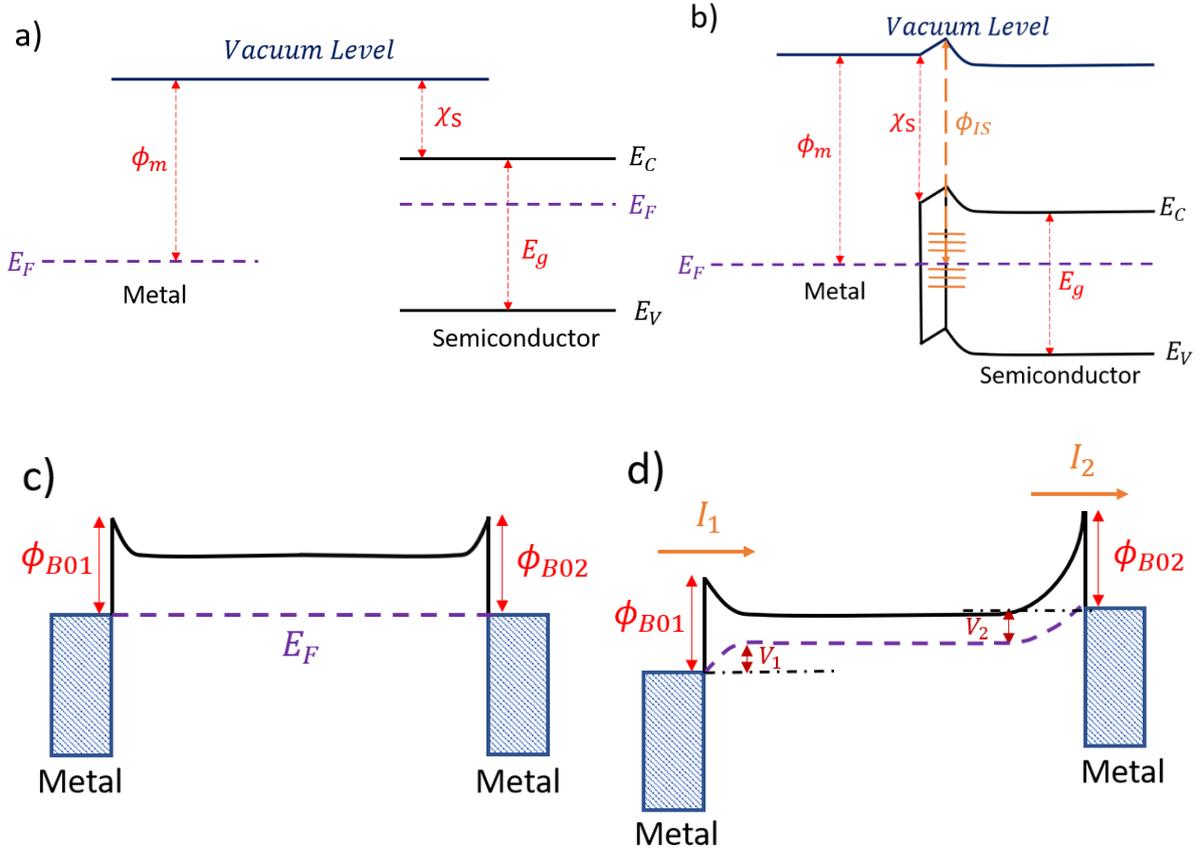

Figure 1 – a) Typical band diagram of a metal and an n-type semiconductor before they are brought into contact. b) Band diagram of a metal in contact with a semiconductor where a high-density of interface states is present. The Fermi level is pinned at the semiconductor interface state, $\phi_{IS}$, and the SB height is independent of the metal work function. c) Energy diagram of a semiconductor with two Schottky junctions at the contacts for unbiased condition. d) Energy diagram of the same device when an external potential V is applied to the left contact. $I_1$ and $I_2$ represent the current through the two barriers, while $V_1$ and $V_2$ are the potential drops at the junctions ($V = V_1 + V_2$). The dashed violet line represents the quasi-Fermi level.

Based on thermionic emission, the current at the two contacts can be written as:[9]

$$I_1 = I_{s1}\left[e^{\frac{qV_1}{kT}} - 1\right] \qquad (3)$$

$$I_2 = -I_{s2}\left[e^{-\frac{qV_2}{kT}} - 1\right] \qquad (4)$$

Where

$$I_{s1,s2} = S_{1,2}\, A^* T^2\, exp\left(-\frac{\phi_{B01,2}}{kT}\right) \qquad (5)$$

are the reverse saturation currents, $A^*$ is the Richardson constant, $T$ is the temperature, $k$ is the Boltzmann constant and $S_{1,2}$ are the areas of the junctions. For the continuity of the current the total current $I_T$ can be expressed as $I_T = I_1 = I_2$, while the applied potential is $V = V_1 + V_2$. Starting from $I_T = I_1$ we have:

$$I_T = I_{s1}\left[e^{\frac{qV_1}{kT}} - 1\right] \quad (6)$$

And substituting $V_1 = V - V_2$:

$$I_T = I_{s1}\left[e^{\frac{qV}{kT}}e^{-\frac{qV_2}{kT}} - 1\right] \quad (7)$$

Using eq. (4), the term $e^{-\frac{qV_2}{kT}}$ can be expressed as

$$e^{-\frac{qV_2}{kT}} = 1 - \frac{I_2}{I_{s2}} = 1 - \frac{I_T}{I_{s2}} \quad (8)$$

since $I_2 = I_T$. Then, substituting eq. (8) into eq. (7), it results:

$$I_T = I_{s1}\left[e^{\frac{qV}{kT}}\left(1 - \frac{I_T}{I_{s2}}\right) - 1\right] \quad (9)$$

Executing all the products and isolating $I_T$ we get:

$$I_T = \frac{I_{s1}\left(e^{\frac{qV}{kT}} - 1\right)}{1 + \frac{I_{s1}}{I_{s2}}e^{\frac{qV}{kT}}} \quad (10)$$

Finally, multiplying numerator and denominator by $2I_{s2}e^{-\frac{qV}{2kT}}$, we obtain

$$I_T = \frac{2I_{s1}I_{s2}\sinh\left(\frac{qV}{2kT}\right)}{(I_{s1}e^{\frac{qV}{2kT}} + I_{s2}e^{-\frac{qV}{2kT}})} \quad (11)$$

As it should be expected, **Equation 11** is transformed into the single SB equation for a configuration with a SB and an ohmic contact, for instance, when $\phi_{B01} \neq \phi_{B02} = 0$. This can be immediately understood considering, for example, that when $\phi_{B02} = 0$ and $S_1 \approx S_2$, it results that

$$I_{s1}e^{\frac{qV}{2kT}} = S_1 A^* T^2 e^{-\frac{\phi_{B01,2}}{kT}}e^{\frac{qV}{2kT}} < I_{s2}e^{-\frac{qV}{2kT}} = S_2 A^* T^2 e^{-\frac{qV}{2kT}} \quad (12)$$

for the low voltages typical of the forward biased Schottky diode ($V < \phi_{B01}$). Therefore, $I_{s1}e^{\frac{qV}{2kT}}$ can be neglected in the denominator of Equation 11, that reduces to the single SB equation:

$$I_T \approx \frac{2I_{s1}I_{s2}\sinh\left(\frac{qV}{2kT}\right)}{I_{s2}e^{-\frac{qV}{2kT}}} = I_{s1}\left(e^{\frac{qV}{kT}} - 1\right) \quad (13)$$

The presence of two equally rectifying junctions is rarely found, especially in small-size devices, where there can be various effects that introduce deviation from the thermionic emission, taken into account by introducing an ideality factor $n$ in the exponential term of **Equation 3** and **Equation 4**. Deviation from the ideality are caused by defects, inadvertent oxide layers and image-force lowering that make the SB height dependent on the applied external voltage. In such a case, the effective SBs $\phi_{B1,B2}(V)$ in **Equation 5** can be written as[9,14]:

$$\phi_{B1,B2}(V) = \phi_{B01,B02} \pm eV_{1,2}\left(1 - \frac{1}{n_{1,2}}\right) \quad (14)$$

Where $\phi_{B01,B02}$ are the ideal SBs at zero bias, $V_1$ and $V_2$ are the voltage drops at the junctions and $n_{1,2}$ are the ideality factors defined as $\frac{1}{n_{1,2}} = 1 \pm \frac{\partial \phi_{B1,B2}}{e\partial V_{1,2}}$.

**Figure 2a** shows I-V curve given by Equation 11 for two junctions with the same reverse saturation current and equal barrier height, at the temperature of 300 K. For both positive and negative voltages, the current saturates at the limit value set by the reverse saturation current of the reverse-biased junction. The reverse saturation current is set by the SB height as well as by junction area, as shown in **Figure 2b**. Indeed, imposing $S_1 = 2S_2$, two saturation currents that are one the double of the other are obtained, yielding an asymmetric I-V characteristic.

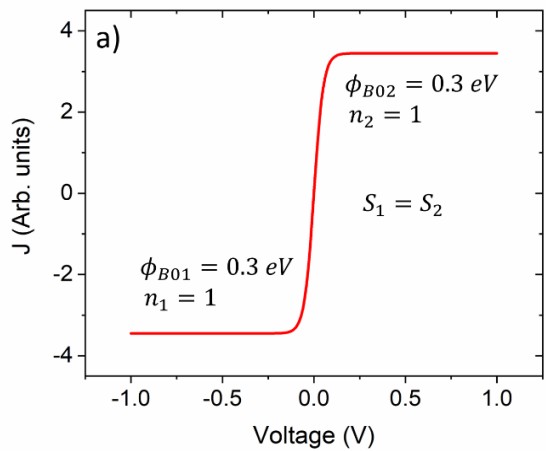
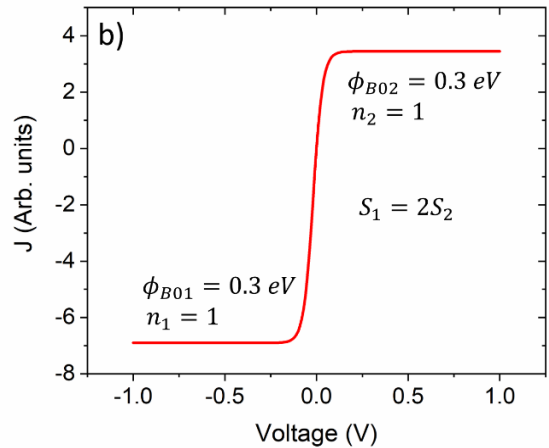
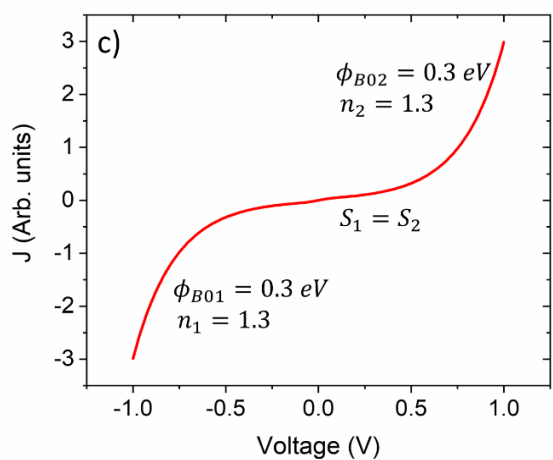
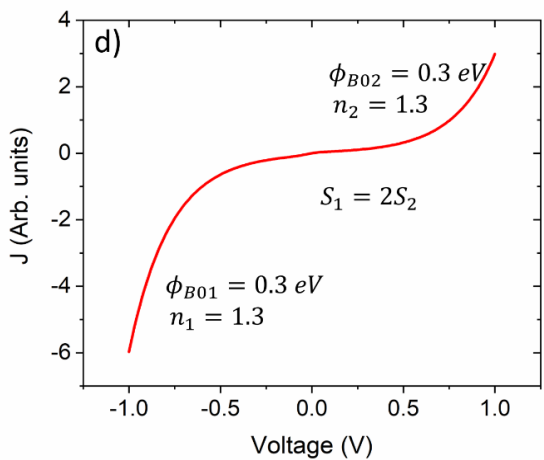

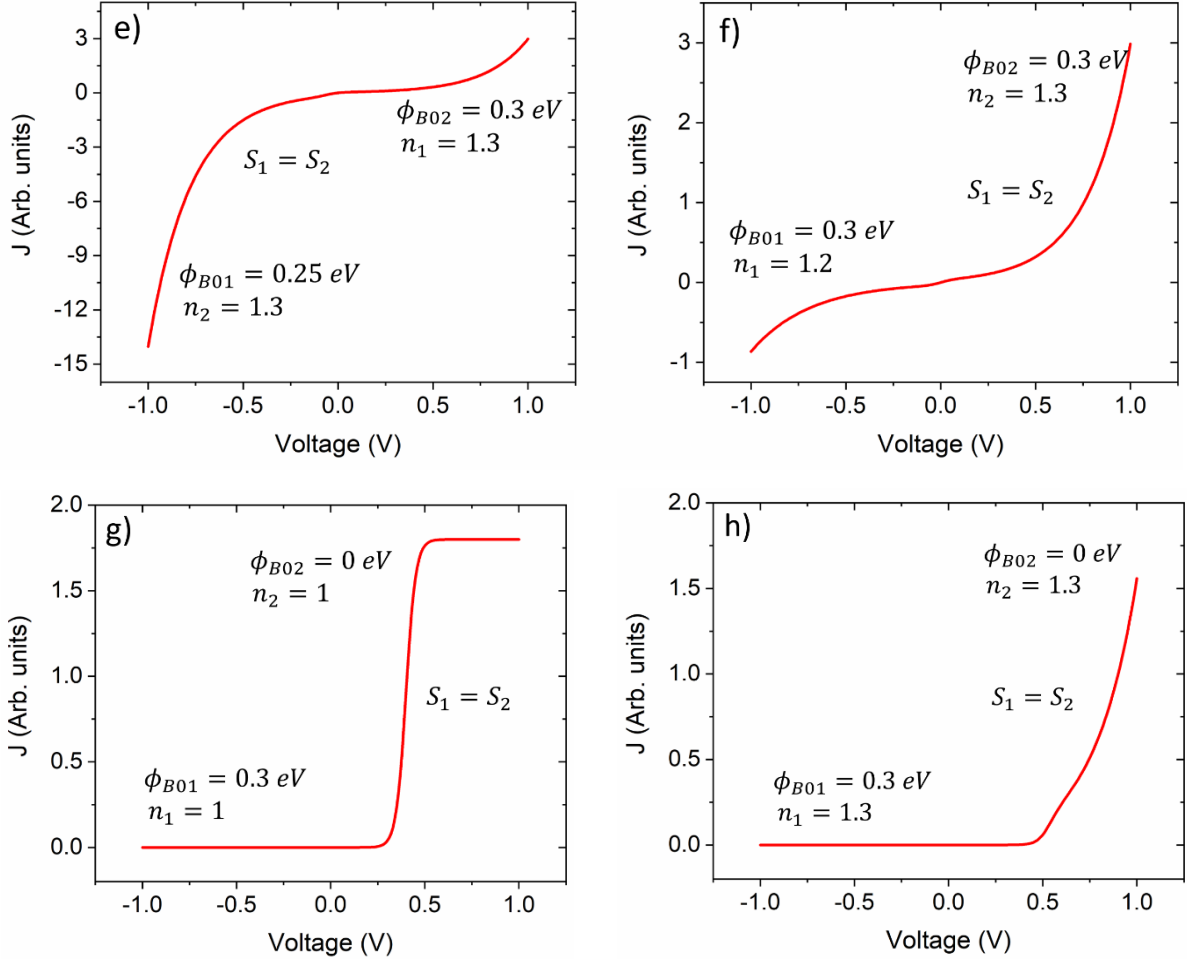

*Figure 2 – a-h) Simulation of eq. (11) describing the current in two back-to-back Schottky junctions. a) I-V characteristic for contacts with the same barrier height, $\phi_{B01} = \phi_{B02} = 0.3\ eV$, the same junction area $S_1 = S_2$ and perfect ideality ($n = 1$). The current saturates to the reverse saturation current of the reverse-biased diode. b) I-V characteristic for $\phi_{B01} = \phi_{B02} = 0.3\ eV$, perfect ideality ($n = 1$), and $S_1 = 2S_2$; the saturated currents are $J_{S2} = 2J_{S1}$. c) I-V characteristic for $\phi_{B01} = \phi_{B02} = 0.3\ eV$, areas $S_1 = S_2$ and ideality factors $n_1 = n_2 = 1.3$. The ideality factor is due to the image force barrier lowering. The current does not saturate because the reverse biased diode is affected by image force barrier lowering. d) I-V characteristic for $\phi_{B01} = \phi_{B02} = 0.3\ eV$, areas $S_1 = 2S_2$ and ideality factors $n_1 = n_2 = 1.3$. We observe that the current is lower when limited by the narrower junction. e) II-V characteristic for $\phi_{B01} = 0.25\ eV$, $\phi_{B02} = 0.3\ eV$, areas $S_1 = S_2$ and ideality factors $n_1 = n_2 = 1.3$. The current results more limited at positive voltage because it is strongly limited by the higher $\phi_{B02}$ barrier. f) I-V characteristic for $\phi_{B01} = \phi_{B02} = 0.3\ eV$, areas $S_1 = S_2$ and ideality factor $n_1 = 1.2$ and $n_2 = 1.3$. The current is limited at negative voltage because of the higher blocking effect of $\phi_{B01}$, that is closer to an ideal barrier. g) I-V characteristic for $\phi_{B01} \neq \phi_{B02} = 0\ eV$, areas $S_1 = S_2$ and perfect ideality ($n_1 = n_2 = 1$). The current of a single ideal Schottky barrier is obtained at low forward bias ($V < \phi_{B01}$). h) I-V characteristic for $\phi_{B01} \neq \phi_{B02} = 0\ eV$, areas $S_1 = S_2$ and $n_1 = n_2 = 1.3$. The negative current is still very low, but the slope of the I-V curve is voltage dependent.*

**Figures 2c-2f** show the plots obtained substituting $\phi_{B1,B2}(V)$ from **Equation 14** in Equation 11 and allowing a slight deviation from the ideal case. In these plots, we have made the simplifying assumption that $V_1 = V_2 = \frac{V}{2}$.

Figure 2c, in particular, shows that despite the reverse-bias of one of the two junctions, the current does not saturate at positive or negative bias due to the dependence of the barrier on the applied voltage. Such behaviour is much more similar to what is usually observed in practical devices. Furthermore, if we consider also different junction areas, the current can increase with different trends, as shown in Figure 2d. In particular, the current is more limited by the junction with the narrower area. Figures 2e and 2f show the I-V curves from Equation 11 with different SB heights and ideality factors. We note that in Figure 2e the current is lower at positive voltage because the current is limited by the higher $\phi_{B02}$ barrier (in contrast, at negative voltage the current is limited by the lower $\phi_{B01}$ barrier). Figure 2f shows the current behaviour in a device where the two Schottky junctions have the same barrier heights but different ideality factors. In this case, the current is lower when is limited by the junction with the lower ideality factor, i.e by the barrier closer to the ideal SB.

**Figure 2g** and **Figure 2h** show I-V curves given by Equation 11 setting $\phi_{B01} \neq \phi_{B02} = 0$. The classic current behaviour for a single Schottky barrier is obtained for both ideal and image-force lowered barriers.

## 3. Results and discussion

In recent years, nanostructured materials have attracted significant attention owing to their unique electrical, optical and mechanical properties.[15–19] Mono and two-dimensional (1D and 2D) nanomaterials present atomic dimensions that enable aggressive length scaling in the future generation of electronic devices like field effect transistors[20–23] and memory devices.[24–27] Despite the huge progresses, the realization of reliable contacts with nanostructured materials poses substantial issues[28–30] that hamper the understanding of the intrinsic electric transport properties and the exploitation of the nanomaterials in electronic and optoelectronic applications. Contacts are the communication links between the nanomaterials and the outer world, and the fabrication of ohmic contacts with linear current-voltage characteristics and low-resistance is still an open challenge.[30–32] One of the main problems is the lack of reproducibility as contacts realized under the same conditions and using the same metals can have different electrical characteristics ranging from Schottky to ohmic behaviour.[33] Moreover, the contacts can influence the conduction mechanism inside a device contributing to the change from the typical thermally activated band conduction to variable range hopping or space charge limited current.[34–36]

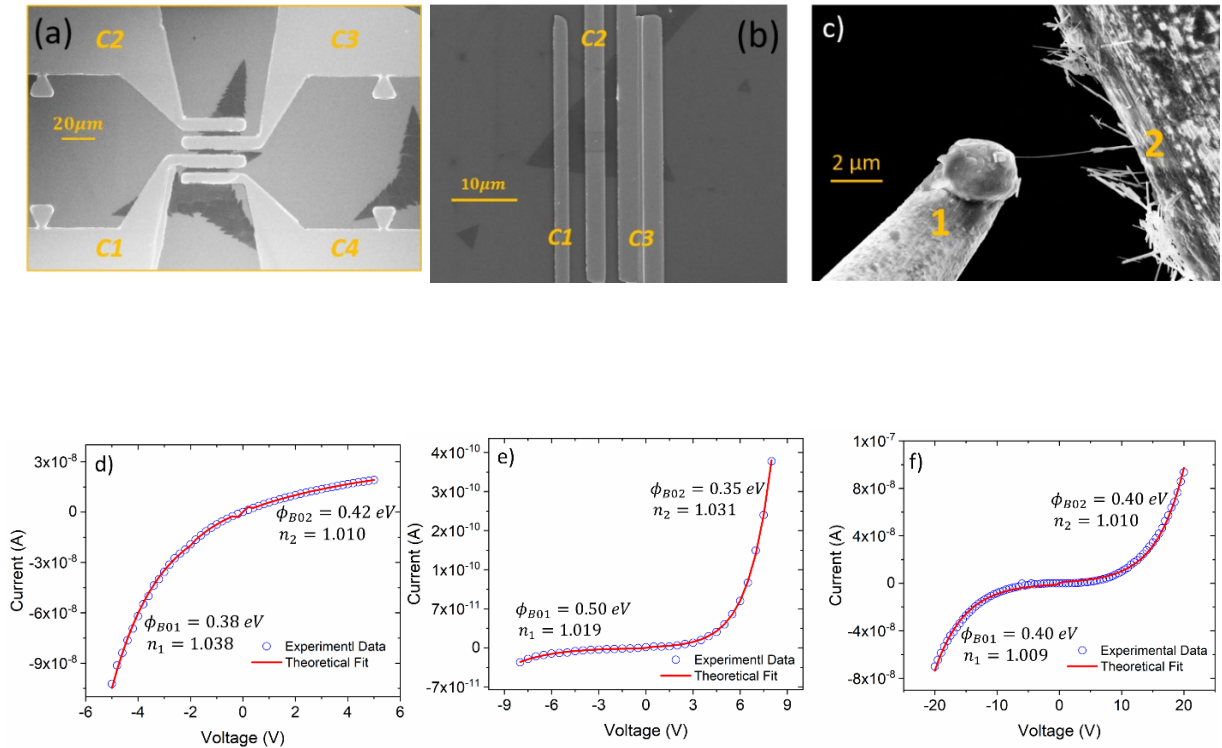

Figure 3 – a) SEM image of a MoS$_2$ flake contacted with four Ti/Au leads. b) SEM image of a WSe$_2$ flake covered with Ni/Au contacts. c) SEM image showing a single WS$_2$ nanotube contacted by two tungsten tips, marked as 1 and 2 respectively. d)-e)-f) I-V characteristic of the MoS$_2$, WSe$_2$ and WS$_2$ devices. The blue circles represent the experimental data, while the red lines are the fitting curves from equation 11, with the SB height and the ideality factor as fitting parameters.

**Figure 3a-c** show the scanning electron microscope (SEM) images of three nanodevices: the first is a MoS$_2$ flake contacted with four Ti/Au leads,[37] the second is a WSe$_2$ flake covered with Ni/Au contacts[23] and the third is a single WS$_2$ nanotube stuck to two tungsten tips due to van der Waals interaction.[38] MoS$_2$ and WSe$_2$ flakes lay on an electrically grounded SiO$_2$/p-Si substrate, while WS$_2$ nanotube is suspended between the two tungsten tips. The details of the fabrication, characterization and electrical measurements can be found elsewhere.[23,37,38]

**Figure 3d** shows the I-V characteristic of the MoS$_2$ nanosheet with strongly asymmetric behaviour. This result confirms that contacts made simultaneously and under the same conditions can have different barrier height (the contact area is about the same). Figure 3d also shows the curve obtained from Equation 11, superposed to the experimental data, demonstrating the excellent fitting. The results of the fit confirm the presence of two different barriers at the contacts that occur to be $\phi_{B01} = 0.38\ eV$ and $\phi_{B02} = 0.42\ eV$ with ideality factor $n_1 = 1.038$ and $n_2 = 1.010$ respectively. From the literature the MoS$_2$ electron affinity ranges from $3.74\ eV$ to $4.45\ eV$,[39–41] while Ti work function is $4.33\ eV$. According to Equation 1, she SB height should range from $-0.12\ eV$ to $0.59\ eV$ where the negative values can be considered as the ohmic contacts. We highlight how equation 1, also known as the Schottky-Mott rule, does not yield an accurate estimate of the barrier, especially for two-dimensional materials. Indeed, the Schottky-Mott model does not take into account the pinning of the Fermi level which may occur at both interfaces, even in an asymmetrical way. For this reason, it can only be used as a qualitative method to obtain the rough estimation of the SB. Moreover, our results are comparable with the SB height ranging from $0.31\ eV$ to $0.51\ eV$ experimentally determined in previous works.[6]

**Figure 3e** shows the I-V characteristic of the WSe$_2$ flake that presents a strong asymmetric behaviour. As expected, the fit yields two different values for the SB heights, namely $\phi_{B01} = 0.50\ eV$ and $\phi_{B02} = 0.35\ eV$ with ideality factor $n_1 = 1.019$ and $n_2 = 1.031$ respectively. In this case, considering that the Ni work function is about $5.0\ eV$ while the WSe$_2$ electron affinity is about $3.7\ eV$[42], the results from the fit of Equation 11 is not consistent with of Equation 1, that would give the unrealistic high barrier of $1.3\ eV$.

**Figure 3f** shows the I-V characteristic of the nanowire of Figure 3c. The I-V curve is symmetric pointing to the presence of two equal barriers. The fit of Equation 11 perfectly reproduces the experimental data and allows the estimation of the SB height as $\phi_{B01} = \phi_{B02} = 0.40\ eV$ with
12

ideality factor $n_1 = 1.009$ and $n_2 = 1.010$ respectively. In this case considering $\chi_s(WS_2) = 4.5\ eV$[43] and $\phi_M(W) = 4.32\ to\ 5.22\ eV$, the estimated barrier is within the range predicted by Equation 1, that is $-0.18\ eV$ to $0.72\ eV$.

**4. Conclusion**

In conclusion, we have proposed a simple model to describe the I-V characteristics of a generic semiconductor device with two Schottky junctions at the contacts. The model includes non-idealities that make the barrier height dependent on the applied voltage. We have shown that the model can account for several experimental observation and fully reproduce both symmetric and asymmetric characteristics. Finally, we have tested the model by fitting the I-V characteristics of devices with MoS$_2$ and WSe$_2$ nanosheets and WS$_2$ nanotubes. The model not only perfectly reproduces the experimental data but also allows to simultaneously extract the two contact barriers and their respective ideality factors.

**Acknowledgements**
((Acknowledgements, general annotations, funding. Other references to the title/authors can also appear here, such as "Author 1 and Author 2 contributed equally to this work."))

Received: ((will be filled in by the editorial staff))
Revised: ((will be filled in by the editorial staff))
Published online: ((will be filled in by the editorial staff))